\documentstyle[float,epsf,12pt]{article}
\makeatletter
\ifcase\@ptsize
 \font\teneufm=eufm10
 \font\seveneufm=eufm7
 \font\fiveeufm=eufm5
 \font\teneusm=eusm10
 \font\seveneusm=eusm7
 \font\fiveeusm=eusm5
\or
 \font\teneufm=eufm10 scaled \magstephalf
 \font\seveneufm=eufm7
 \font\fiveeufm=eufm5
 \font\teneusm=eusm10 scaled \magstephalf
 \font\seveneusm=eusm7
 \font\fiveeusm=eusm5
\or
 \font\teneufm=eufm10 scaled \magstep1
 \font\seveneufm=eufm7
 \font\fiveeufm=eufm5
 \font\teneusm=eusm10 scaled \magstep1
 \font\seveneusm=eusm7
 \font\fiveeusm=eusm5
\fi

\newfam\eufmfam
\newfam\eusmfam
\textfont\eufmfam=\teneufm  \scriptfont\eufmfam=\seveneufm
  \scriptscriptfont\eufmfam=\fiveeufm
\textfont\eusmfam=\teneusm  \scriptfont\eusmfam=\seveneusm
  \scriptscriptfont\eusmfam=\fiveeusm

\def\frak{\ifmmode\let\next\frak@\else
 \def\next{\errmessage{Use \string\frak\space only in math mode}}\fi\next}
\def\frak@#1{{\frak@@{#1}}}
\def\frak@@#1{\fam\eufmfam#1}

\def\sh{\ifmmode\let\next\sh@\else
 \def\next{\errmessage{Use \string\sh\space only in math mode}}\fi\next}
\def\sh@#1{{\sh@@{#1}}}
\def\sh@@#1{\fam\eusmfam#1}

\ifcase\@ptsize
 \font\tenmsa=msam10
 \font\sevenmsa=msam7
 \font\fivemsa=msam5
 \font\tenmsb=msbm10
 \font\sevenmsb=msbm7
 \font\fivemsb=msbm5
\or
 \font\tenmsa=msam10 scaled \magstephalf
 \font\sevenmsa=msam7
 \font\fivemsa=msam5
 \font\tenmsb=msbm10 scaled \magstephalf
 \font\sevenmsb=msbm7
 \font\fivemsb=msbm5
\or
 \font\tenmsa=msam10 scaled \magstep1
 \font\sevenmsa=msam7
 \font\fivemsa=msam5
 \font\tenmsb=msbm10 scaled \magstep1
 \font\sevenmsb=msbm7
 \font\fivemsb=msbm5
\fi

\newfam\msafam
\newfam\msbfam
\textfont\msafam=\tenmsa  \scriptfont\msafam=\sevenmsa
  \scriptscriptfont\msafam=\fivemsa
\textfont\msbfam=\tenmsb  \scriptfont\msbfam=\sevenmsb
  \scriptscriptfont\msbfam=\fivemsb

\def\Bbb{\ifmmode\let\next\Bbb@\else
 \def\next{\errmessage{Use \string\Bbb\space only in math mode}}\fi\next}
\def\Bbb@#1{{\Bbb@@{#1}}}
\def\Bbb@@#1{\fam\msbfam#1}
\def\hexnumber@#1{\ifnum#1<10 \number#1\else
 \ifnum#1=10 A\else\ifnum#1=11 B\else\ifnum#1=12 C\else
 \ifnum#1=13 D\else\ifnum#1=14 E\else\ifnum#1=15 F\fi\fi\fi\fi\fi\fi\fi}
\def\msa@{\hexnumber@\msafam}
\def\msb@{\hexnumber@\msbfam}
\mathchardef\square="0\msa@03

\makeatother
\newcommand{\beq}{\begin{equation}}
\newcommand{\eeq}{\end{equation}}
\newcommand{\ba}{\begin{array}}
\newcommand{\ea}{\end{array}}
\newcommand{\bea}{\begin{eqnarray}}
\newcommand{\eea}{\end{eqnarray}}
\newcommand{\bean}{\begin{eqnarray*}}
\newcommand{\eean}{\end{eqnarray*}}

\newtheorem{theorem}{Theorem}[section]
\newtheorem{prop}[theorem]{Proposition}

\newtheorem{defi}[theorem]{Definition}
\newtheorem{remark}[theorem]{Remark}

\newtheorem{proof}{Proof.}
\newenvironment{pf}{\begin{proof} \rm}{\hfill$\square$ \end{proof}}
\newcommand{\CH}{{\cal H}}
\newcommand{\CL}{{\cal L}}

\newcommand{\CW}{{\cal W}}
\newcommand{\CV}{{\cal V}}
\newcommand{\CU}{{\cal U}}
\newcommand{\CZ}{{\cal Z}}
\newcommand{\CS}{{\cal S}}
\newcommand{\CN}{{\cal N}}
\newcommand{\CM}{{\cal M}}
\newcommand{\CQ}{{\cal Q}}
\makeatletter
\@addtoreset{equation}{section}

\makeatother

\def\be{\beta}
\def\al{\alpha}

\def\la{\lambda}

\newcommand{\cmp}[3]{Comm. Math. Phys. {\bf #1} (#2), #3}

\newcommand{\faa}[3]{Funct. Anal. Appl. {\bf #1} (#2), #3}

\newcommand{\ijmp}[3]{Int. Jour. Mod. Phys. A{\bf #1} (#2), #3}

\newcommand{\lmp}[3]{Lett. Math. Phys. {\bf #1} (#2), #3}

\newcommand{\mplA}[3]{Mod. Phys. Lett. {\bf A#1} (#2), #3}

\newcommand{\rref}[1]{(\ref{#1})} 

\newcommand{\del}{{\partial}}

\def\Fdb{{Fa\`a di Bruno}}
 
\def\dpt#1#2{\frac{\partial #1}{\partial t_{#2}}} 
\def\H#1{H^{(#1)}} 
\def\h#1{h^{(#1)}}

\def\F#1{F^{(#1)}}

\def\v#1{{v^{(#1)}}}
\def\V#1{{\CV^{(#1)}}}
\def\endpf{\begin{flushright}$\square$\end{flushright}}
\def\kp#1{{KP$^{{(#1)}}$}}
\def\cs#1{{CS$_{{#1}}$}}
\def\kdv#1#2{{KdV$^{{#1}}_#2$}}

\def\CHp{{\CH_+}}
\def\CHm{{\CH_-}}
\def\Alg{{\frak G}}                                
                              
\def\alg{{\frak g}}
\newcommand{\fraksl}{{\frak s}{\frak l}}
\def\ger{hierarch}
\def\var{manifold}

\def\bih{bihamiltonian}
\def\varb{\bih\ \var}

\def\parp{Poisson bracket}                    
\def\ger{hierarch}

\def\tenp{Poisson tensor} 
\def\varb{\bih\ \var}
\def\res{\mbox res}
\def\qmn#1#2{{\CQ^{#1}_{#2}}}
\def\sl23{{\fraksl}_3^{(2)}} 
\begin{document}
\begin{titlepage}
\begin{flushright}
Ref. SISSA 94/96/FM
\end{flushright}
\begin{center}
{\huge 
A Note on Fractional KdV Hierarchies}
\end{center}
\begin{center}
{\large
Paolo Casati${}^1$, Gregorio Falqui${}^2$,\\
Franco Magri${}^1$, and
Marco Pedroni${}^3$}\\
${}^1$ Dipartimento di Matematica, Universit\`a di Milano\\
Via C. Saldini 50, I-20133 Milano, Italy\\
E--mail: casati@vmimat.mat.unimi.it,
magri@vmimat.mat.unimi.it\\ 
${}^2$ SISSA/ISAS, Via Beirut 2/4, I-34014 Trieste, Italy\\
E--mail: falqui@sissa.it\\ 
${}^3$ Dipartimento di Matematica, Universit\`a di Genova\\
Via Dodecaneso 35, I-16146 Genova, Italy\\
E--mail: pedroni@dima.unige.it
\end{center}
\abstract{\noindent
One of the cornerstone of the theory of integrable systems of KdV type has been the remark that the $n$--GD equations are reductions of the full KP theory.
In this paper we address the analogous problem for the fractional 
KdV theories, by suggesting candidates of the ``KP--theories'' lying behind 
them. These equations are called ``\kp{m} hierarchies'', and are obtained as 
reductions of a bigger dynamical system, which we call the ``central system''.
The procedure allowing to pass from the central system to the \kp{m} equations,
and then to the fractional \kdv{m}{n} equations, is discussed 
in detail in the paper.  
The case of  \kdv{2}{3} is given as paradigmatic example.}
\\
\\
Work supported by the Italian M.U.R.S.T. and by
the G.N.F.M. of the Italian C.N.R.\\
One of the authors (F.M.) thanks the organizers of  
the {\em Semester on Integrable Systems}
for the kind hospitality at the {\em Centre 
\'Emile Borel}, where part of this work has been
completed 
\end{titlepage}
\setcounter{footnote}{0}

\section{Introduction}\label{sec1}
The study of integrable systems of KdV type
received, in the last few years, a new impulse from
important developments in  Two--Dimensional (Quantum) Field Theory.
In this framework, much attention has been paid to the conformal 
$\CW_N$--algebras of symmetries of these theories.
In the cases studied at first, they have been shown~\cite{FaLu90,Kh87}
to be quantum extension
of the second Poisson structure of the  $n$--GD (Gel'fand Dickey)
theories~\cite{DS,Dik} associated with a classical Lie algebra ${\frak g}$.
The powerful method of Hamiltonian reduction has been widely 
applied to study and classify those hierarchies of  
partial differential equations and the associated $\CW$--algebras
(see, e.g., \cite{W-rev}).
A class of new integrable hierarchies 
(called fractional KdV or generalized DS)
has been obtained~\cite{Be91,BaDe91,Hol1,Hol2,DeMa92,FeHaMa93} 
by means of a generalization of the 
Drinfel'd--Sokolov construction.
Roughly speaking, fractional KdV \ger ies correspond 
to the case in which the value of the momentum mapping for the
infinitesimal gauge action of the loop algebra $L \alg$ 
is different from the sum of the (duals of the) 
simple positive roots of $\alg$. They have been
shown~\cite{Hol1,Hol2,FGMS95} 
to be classified by homogeneous elements in 
Heisenberg subalgebras 
of the affinization $\tilde \alg$ of $\alg$.\par
One of the cornerstones of the theory 
of integrable systems of KdV type 
is the fact that $n$--GD theories can be obtained from the
full Kadomtsev--Petviashvilij (KP) hierarchy
by means of a suitable reduction process~\cite{Dik,SW,DJKM}. 
In this paper we study the problem of the generalization
of such a link to the case of fractional KdV theories.
Namely we want to suggest candidates of ``KP--theories''
lying behind fractional KdV ones.\par
Our starting point is a special hierarchy of dynamical systems 
described by equations of Riccati type, which we call the {\em central 
system (CS).}
They are defined on the space $\CH$ of sequences  
$\{\H{k}\}_{k\in\Bbb N}$
of Laurent series of the form
\beq
\label{h3a}
\H{0}=1,\qquad
H^{(k)}=z^k+\sum_{l\ge 1}H^k_l z^{-l}\\
\end{equation}
by the following equations
\beq
\label{00a}
\frac{\partial H^{(k)}}{\partial t_j}+H^{(j)}H^{(k)}=H^{(j+k)}+\sum_{l=
1}^kH^j_lH^{(k-l)}+
\sum_{l=1}^jH^k_lH^{(j-l)}.
\eeq
They  arise
in the framework of the \bih\ approach to the KdV equation, as we
shall see in Section~\ref{sec2}. 
One can remark that the space $\CH$ is a subspace of the big cell
of the 
Sato Grassmannian $Gr(H)$~\cite{SS,SW}, and therefore, one can suspect that 
these equations are related to the linear flows of 
the $gl_\infty$--action on $Gr(H)$. This is indeed correct, since one can prove that the two types of equations are related by a Darboux transformations. The 
study of this connection, however, is outside the scopes of the present paper 
~\cite{FMP}. 
Our aim, in this paper, is to  derive the fractional  KdV
equations from CS by a double process of
reduction.\par
The ideas of the present approach are conveniently described
by adopting a 
geometric point of view. 
We regard equations~\rref{00a} as 
defining a hierarchy of vector fields $X_j$ on $\CH$. 
We notice that these vector fields commute, and we consider the space
\bea
\CQ^m=\CH/X_m\label{spq1}
\eea
of the orbits of the vector field $X_m$. 
Since the CS flows commute, each vector field 
$X_j$ sends solutions of $X_m$ into 
different solutions, and therefore induces a flow on the space
$\CQ_m$. 
We shall call the hierarchy of the reduced flows 
on $\CQ_m$ the {\em \kp{m} hierarchy}.
The reason is that for $m=1$, one obtains in this 
way the usual KP hierarchy in the $1+1$--dimensional picture.
We were led to consider different values of $m$ by 
the conjecture~\cite{DeMa92,OlMa91} 
that $m$--fractional KdV \ger ies can be obtained
by ``exchanging the roles of $t_m$ and $t_1$'' in the $m$--GD
equations. 
However, we stress that the projection onto $\CQ_m$,
which can also be  thought of as a field--theoretic 
redefinition of flows and ``independent variables'',
in our geometrical scheme is more conveniently
considered as a reduction of the central system.
\par
Besides the space $\CQ^m$, we consider  the manifold
\beq
{\CZ}_n :=\{H\in\CH\mid X_n(H)=0\}
\label{defcsn}
\eeq
of the zeroes of $X_n$ and its submanifold 
\beq
\CS_n=\{H\in\CZ_n\mid \H{n}=z^n\},
\eeq
where the Laurent series $\H{n}$ assumes the constant value
$\H{n}=z^n$. We remark that they are invariant 
submanifolds for the central system.
The restrictions of the vector fields $X_j$ to $\CS_n$ 
give rise to a second system of reduced 
equations which we call {\em CS${}_n$ equations.\/}  
Finally,  we combine the two reduction processes, 
by considering the space 
\beq
\CQ_n^m=\CS_n/X_m
\eeq
of the orbits of the vector field $X_m$ restricted to $\CS_n$.
We obtain a reduction 
of the CS${}_n$ equations which we call {\em \kdv{m}{n} equations.}
Since the reduction processes commute, they can also be seen as
restrictions of the \kp{m} equations to $\CQ_n^m$.
The relevant manifolds are shown in 
Figure~1.
\begin{figure}[ht]
  \caption{The Reduced Manifolds}
\bigskip\bigskip
\centerline{\epsfxsize=6.cm\epsfbox{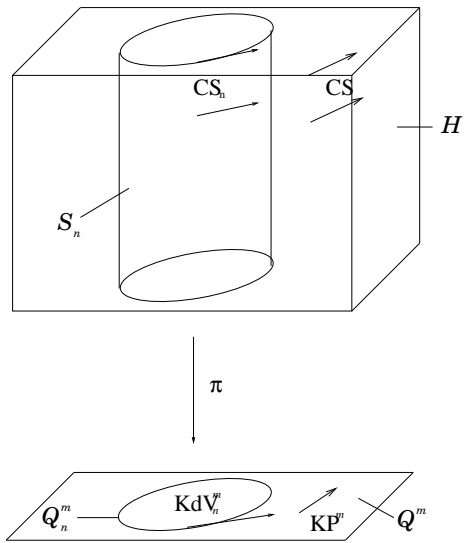}} 
 \end{figure}
\par 
It is our belief that, when $m$ and $n$
are coprime, the \kdv{m}{n} equations  coincide with the fractional KdV 
equations (or generalized type I GD equations) 
studied in~\cite{Hol1,FGMS95}.
This will be shown explicitly in the case of \kdv{2}{3}.\par
Other interesting equations can be obtained by different choices of the 
reduction spaces. So, for example,
the restrictions of $X_j$ to the intersection \hbox{{\CS}$_n\cap$ 
{\CZ}$_p$} of two
invariant manifolds give rise to finite dimensional dynamical
systems of Bogoyavlensky--Dubrovin--Novikov type \cite{BN,DMN}.
Similarly, the projection of the \kp{m} hierarchy into the space 
\beq
\CQ^{(m,l)}={\CQ}^m/X_l\label{cqml}
\eeq
of the orbits of the reduction of $X_l$ to $\CQ^m$, can lead to 
equations in  two space dimensions. They are denoted  \kp{m,l} in 
Figure~2, where  
the full reduction scheme is represented;  
the left arrows denote a restriction to invariant submanifolds, and 
the right arrows denote a projection onto orbit spaces $\CQ_m$.\par
The main aim of this paper is to study the equations
CS${}_n$, \kp{m} and \kdv{m}{n} and their relations.
The paper is organized as 
follows. Section~\ref{sec2} is a brief introduction
to the central system from the point of view of the bihamiltonian
approach to the KdV equations. The first  fundamental 
properties of CS are studied in Section~\ref{sec3}, where
we describe in detail the submanifolds $\CS_n$ 
and the quotient spaces $\CQ^m$. 
Section~\ref{sec4} 
gives a preliminary view of the equations
which can be obtained by iterating and combining
the reduction processes in the simplest case  $m=1$, 
$n=2$ corresponding to the usual KP theory. 
In Section~\ref{sec5} we exhibit the 
explicit structure of the equations \cs{n} and  \kp{m}.
Section~\ref{sec6} is devoted to the \kdv{m}{n} equations. 
In particular we work out the case \kdv{2}{3}. 
Finally, in Section~\ref{sec7} we give the bihamiltonian 
interpretation of these equations along the lines explained in 
Section~\ref{sec2}, and we make an explicit comparison 
with the approach based on the generalized DS procedure.
\begin{figure}[hb]
\begin{center}
\caption{The Full Reduction Schemes}
\end{center}
\setlength{\unitlength}{0.5cm}
\begin{center}
\begin{picture}(14,8)(-7,-4)
\put(0,3.5){\makebox(0,0)[b]{CS}}
\put(0.2,3){\vector(1,-1){3}}
\put(0,3){\vector(-1,-1){3}}
\put(3.2,0){\makebox(0,0)[tl]{KP${}^{(m)}$}}
\put(-3,0){\makebox(0,0)[tr]{CS${}_n$}}
\put(-3.7,-0.7){\vector(-1,-1){3}}
\put(-6.7,-3.7){\makebox(0,0)[tr]{BDN${}_{n,p}$}}
\put(-3,-0.7){\vector(1,-1){3}}
\put(0,-3.7){\makebox(0,0)[t]{KdV${}_n^{(m)}$}}
\put(3.2,-0.7){\vector(-1,-1){3}}
\put(3.9,-0.7){\vector(1,-1){3}}
\put(6.9,-3,7){\makebox(0,0)[tl]{KP${}^{(m,n)}$}}
\end{picture}
\end{center}
\end{figure}
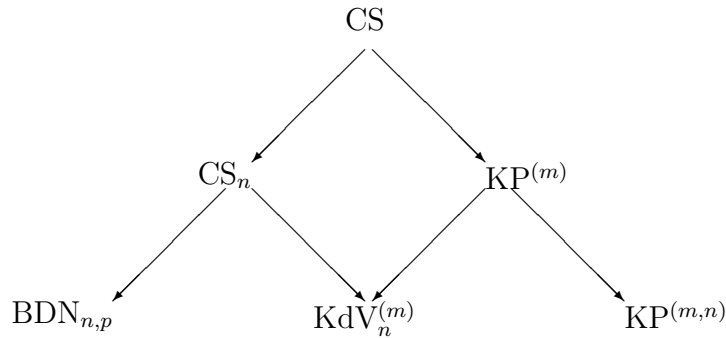

\section{The Central System}\label{sec2}

The present approach to fractional KdV hierarchies comes 
from the bihamiltonian theory of KdV equation. It may be suitable,
therefore, 
to collect in this section the main ideas of this theory.\par
The starting point is the relation between bihamiltonian manifolds and 
integrable systems clarified by a theorem of Gel'fand and Zakharevich 
\cite{GZ}. Let the phase space $\CM$ be a $(2n+1)$--dimensional manifold, 
endowed with a pencil 
of compatible Poisson brackets 
\beq
\{f,g\}_\la=\{f,g\}_1-\la \{f,g\}_0.\label{pbp}
\eeq
We assume that this pencil has maximal rank everywhere 
on $\CM$, and that the symplectic leaves of $\{f,g\}_\la$ are 
submanifolds of dimension $2n$ in $\CM$. 
As it was noticed  by Gel'fand and Zakharevich,
its Casimir 
function is a polynomial of degree $n$ in the parameter $\la$ 
\beq
H(\la)=H_0+_1\la+\dots+H_n{\la}^n,\label{cf}
\eeq
and its coefficients
$(H_0,H_1,\dots,H_n)$ are in involution with respect 
to all the Poisson brackets of the pencil.
Therefore, they verify the conservation laws
\beq
\dpt{}{j}H_k=0,
\eeq
where the derivative of the function $H_k$ is taken along the Hamiltonian
vector field associated
with the function $H_j$.
These vector fields define  
an integrable system on $\CM$.\par
This bihamiltonian strategy may be formally extended to
partial differential equations in one space variable $x$, 
with two significant differences.  
The Casimir functions $H(\la)$ now becomes a 
Laurent series in $\la$ (or in some power of $\la$) and 
the involution relations $\{H_i,H_j\}_\la=0$,  $i,j=0,\dots n$, 
are replaced by local conservations laws (or continuity equations) 
of the form 
\beq 
\frac{\del h}{\del t_j}=\del_x \H{j}.\label{conlaw}
\eeq
This fact is responsible for the appearance of a further 
important object of the theory: the {\em current densities 
$\H{j}$.} They define a new geometrical structure associated with the 
points of the phase space, which evolves in time along the orbits 
of the equations~\rref{conlaw}. The central system is the system
of equations describing the time evolution of the current densities $\H{j}$.
\par
Let us  describe these features in the example of the KdV 
equations. From the bihamiltonian point of view, 
the KdV theory may be seen as 
the  study of the Casimir function and the current densities 
associated with the Poisson pencil 
\beq
{\dot u}=-\frac12 v_{xxx}+2(u+\la)v_x+u_xv.\label{rp}
\eeq
This pencil can be  obtained by a 
Marsden--Ratiu reduction of a Lie--Poisson pencil  
defined over the ${\frak s}{\frak l} (2)$ loop algebra
\cite{CMP,CFMP23}. Here, $u$ is a point of the phase space of the KdV theory, 
$v$ is a covector, and the Poisson pencil is a map from the cotangent space 
to the tangent space.\par  
The computation of the Casimir function is comparatively easy. 
Let us set $\la=z^2$, and let us introduce  
 the Hamiltonian density $h(z)$ 
of the 
Casimir function according to 
\beq
H(z)=2z\int_{S^1}h(z)dx.\label{h1}
\eeq
One can prove that $h(z)$ is the unique monic Laurent
series 
\beq
h(z)=z+\sum_{l\ge1}h_l(x)z^{-l}\label{h2}
\eeq
which solves the Riccati equation
\beq
h_x+h^2=u+z^2.\label{ric}
\eeq
Thus, the role of this equation is  to define the Casimir function at 
any point $u$ of the phase space.\par
The computation of the current $\H{j}$ is, instead, a little more 
tricky. We have to consider the Fa\`a di Bruno polynomials
$\h{k}$ defined  by the recursive 
formula
\beq
\h{k+1}=\h{k}_x+h\h{k},\label{h3}
\eeq
starting from $\h{0}=1$.
Then we have to linearly combine these polynomials,
with coefficients depending on the coefficients $h_i$ of $h(z)$, 
but {\em independent} of $z$,
\beq
\H{j}=\sum_{k=0}^jc^j_k(h_1,h_2,\dots)\h{k}(z)\label{hp}
\eeq
in such a way to obtain Laurent series $\H{j}$
of the form 
\beq
\H{i}(z)=z^j+\sum_{l\ge1}H^i_lz^{-l}.\label{h4}
\eeq
They are the current densities we are looking for.
Indeed one can prove~\cite{Wil81,CFMP23} that, 
when $u$ evolves according to the KdV hierarchy, the
solution $h(z)$ of the Riccati equation~\rref{ric}
evolves exactly according to the conservation laws~\rref{conlaw},
with the currents~\rref{h4}.
Furthermore,
 the current densities presently 
defined evolve in time according to the central system~\rref{00a}.\par
To arrive at the general form of the central system~\rref{00a}, 
we have to remark 
that  the assumption that $h(z)$ be a
solution of the Riccati equation is actually inessential. 
Indeed, equations~\rref{hp} and~\rref{h4} define the current densities $\H{j}$
associated with {\em any} Laurent series $h(z)$ of the form~\rref{h2}. 
Therefore, we can regard equations 
\rref{conlaw} as a system of local conservation laws for any 
monic Laurent series $h(z)$, whether or not it satisfies the Riccati 
equation~\rref{ric}. This enlarged hierarchy 
 is the KP theory. 
Indeed, equations~\rref{conlaw}, considered
on an arbitrary monic Laurent series $h(z)$, 
are (a possible form of) the celebrated KP 
equations \cite{Wil81,SS,CFMP4}. \par
The second remark is that also the definition~\rref{hp} of the 
current densities  we have used to construct the KP equations, 
is inessential.
The central system
is actually independent of the definition~\rref{hp} of the currents
$\H{j}$, and only rests on the particular form of their 
Laurent expansion~\rref{h4}.
Therefore, we can eliminate 
any reference to a ``space variable $x$'' and to a ``Hamiltonian 
density $h(z)$'', and we can regard the central system as defining an 
independent family of vector fields on the space $\CH$
of collections of such currents. 
From this perspective, equations \rref{hp} (associated with 
the KP equations) and the Riccati equation \rref{ric} 
(leading to the KdV theory)  
are  simply  a set of {\em constraints}, 
which are compatible with the central system. 
Other constraints are possible as 
well, leading to different systems of equations. Among them
 there are the  fractional KdV systems. 
\section{The Central System and its Reductions} 
\label{sec3}
In this section we begin to study the central systems
and its reductions. 
The starting point is the observation that the CS vector fields 
pairwise commute. To prove this property, we begin with
some remarks on its structure.\par
Let us consider the space $\CL$ of truncated Laurent series
\beq
l(z)=\sum_{-\infty}^{N(l)}l_jz^j.\label{vvv}
\eeq
In this space the currents $\H{j}$, $j\ge 0$, 
introduced in Section~\ref{sec1} as
\begin{equation}
\H{0}=1,\qquad
H^{(k)}=z^k+\sum_{l\ge 1}H^k_l z^{-l},
\end{equation}
determine a subspace
\beq
\CHp:=\langle\H{0},\H{1},\dots\rangle,
\eeq
transversal to the subspace
\beq
\label{declau}
\CHm:=\langle z^{-1},z^{-2},\dots\rangle
\eeq
of  the Laurent series of strictly negative degree. 
We can now regard any solution of CS as a subspace 
$\CHp(t)$ moving in $\CL$. The characteristic property of the 
solutions of CS is the invariance of $\CHp(t)$ with respect to the 
action of the differential operators $\dpt{}{j}+\H{j}$ associated
with the currents.
\begin{prop}\label{invprop}
Along the flows of the central system the subspace $\CHp(t)$ satisfies 
the invariance relation
\beq
\label{invrel}
\left(\dpt{}{j}+\H{j}\right)(\CHp)\subset\CHp.
\eeq
\end{prop}
{\bf Proof.} We show that this relation completely defines CS. 
Since the currents $\H{k}$ form a basis of $\CHp$, the invariance 
property \rref{invrel} entails the existence of coefficients $c^{jk}_{l}$ 
independent of $z$ such that
\beq
\label{siscenfun}
\dpt{\H{k}}{j}+\H{j}\H{k}=\sum_{l=0}^{j+k}c_l^{jk}\H{l}.
\eeq
These coefficients are easily determined by comparing the 
expansion in powers of $z$ of both sides of 
equation~\rref{siscenfun}.
The final result is equation~\rref{00a} defining the central system.
\endpf
As an immediate consequence of the previous result, we obtain the 
following compact representation of CS. 
We denote by $\pi_+$ and $\pi_-$ 
the canonical projections associated with the decomposition
\beq
\label{plmi}
\CL=\CH_+ \oplus\CH_-
\eeq 
of the space $\CL$. Then we have
\begin{prop}
\label{propforcs}
The central system is the family of dynamical systems on the currents 
$\H{j}$ given by
\beq
\label{forSC}
\dpt{}{k}\H{j}= - \pi_-(\H{j} \H{k}).
\eeq
\end{prop}
{\bf Proof.} We apply $\pi_-$ to both sides 
of~\rref{siscenfun} and we get
\[
\pi_-\left(\dpt{\H{k}}{j}+\H{j}\H{k}\right)=0.
\]
The assertion follows from the fact that
$H^{(k)}=z^k+\sum_{l\ge 1}H^k_l z^{-l}$ so that $(\dpt{\H{k}}{j}) 
\in {\cal H}_-$.
\endpf
From the symmetry in $j$ and $k$ of the right--hand side of~\rref{forSC}, 
we finally 
obtain the following
\begin{prop} 
Every solution of CS satisfies the exactness condition
\beq
\label{j-k}
\dpt{}{k}\H{j}=\dpt{}{j}\H{k}.
\eeq
\end{prop}
\endpf
We are now ready to prove the commutativity property of the CS flows.
\begin{prop}
The flows of the central system pairwise commute.
\end{prop}
{\bf Proof.}
We compute the action of the commutator $[X_j,X_k]$ of two vector fields 
of the hierarchy on a generic current:
\beq
[X_j,X_k](\H{i})=\dpt{}{j}\dpt{\H{i}}{k}-\dpt{}{k}\dpt{\H{i}}{j}.
\eeq
We remark that that this quantity belongs to $\CHm$, 
thanks to the specific form of $\H{j}$. 
Then we observe that, using 
the exactness property \rref{j-k}, this commutator can be also written 
in the form
\[
[X_j,X_k](\H{i})=\left[ \dpt{}{j}+ \H{j}, \dpt{}{k}+\H{k}\right] 
\H{i},
\]
so that the commutator of the vector fields coincides with the 
commutator of the operators $\dpt{}{j}+\H{j}$ associated with the 
currents. Hence the invariance property \rref{invrel} entails that 
$[X_j,X_k](\H{i})$ belongs to the subspace $\CHp$ for all the $\H{i}$'s. 
But $\CHp\cap\CHm
=\{0\}$, and therefore $[X_j,X_k]$ vanishes.
\endpf
\subsection{The Invariant Submanifolds $\CS_n$}
From the commutativity of the flows it follows that the 
set $\CZ_n$ of zeroes of the vector field $X_n$,
defined by the 
quadratic equations
\beq\label{zerixn}
H^{(k+n)}-H^{(k)}H^{(n)}+\sum_{l=1}^kH^n_lH^{(k-l)}+
\sum_{l=1}^nH^k_lH^{(n-l)}=0,
\eeq
is an invariant submanifold for CS. 
Moreover, on $\CZ_n$ we have
\beq\label{0cu}
\dpt{\H{n}}{j}=\dpt{\H{j}}{n}=0,
\eeq
thanks to the exactness property \rref{j-k}.
Therefore the manifold $\CZ_n$ is foliated by invariant 
submanifolds defined by the equation $\H{n}=\mbox{constant}$. 
Among all these leaves we choose the one given by the condition
\beq
\H{n}=z^n,\label{snc}
\eeq
which is the counterpart of the choice usually considered
in the reduction theory from KP to the n--GD hierarchies~\cite{SW,DJKM}.
\begin{defi} We call  $\CS_n$ the submanifold of the zeroes of $X_n$, 
where the current $\H{n}$ satisfies the relation \rref{snc}. 
\end{defi}
The equations defining  this 
submanifold can be explicitly found by eliminating $\H{n}$
from~\rref{zerixn} by means of the constraints~\rref{snc}.
This leads to the conclusion 
that $\CS_n$ is the subset of $\CH$  
defined by the equations
\beq
\H{j+n}=z^n\H{j}-\sum_{l=1}^nH^j_l\H{n-l}.\label{receq}
\eeq
Therefore, it is
parametrized by the first $n-1$ currents 
$(\H{1},\dots,\H{n-1})$.\\
A more intrinsic characterization is provided by the following 
\begin{prop} 
\label{propsn}
The submanifold $\CS_n$ is the subset of $\CH$  
given by the equation
\beq
z^n(\CHp)\subset \CHp,\label{zinv}
\eeq
i.e., the set of the points where the operator 
of multiplication by $z^n$ leaves the space $\CHp$ invariant.
\end{prop}
{\bf Proof.}  First of all, it is easily shown that the condition 
\rref{zinv} is necessary. It is enough, indeed, to observe that the 
invariance relation 
\beq
\left(\dpt{}{n}+\H{n}\right)(\CHp)\subset \CHp
\eeq
characterizing the vector field $X_n$, at the points of $\CS_n$ 
boils down to the relation \rref{zinv} thanks to~\rref{snc}. 
\\
To show that the condition is also sufficient, let us remark 
that $z^n\cdot \H{0}=z^n\cdot 1=z^n$ belongs to $\CHp$ and then can be 
developed as
\beq
z^n=\sum_{l=0}^n c^n_l\H{l}.
\eeq
By a comparison between the coefficients of the positive powers of $z$ 
we obtain that
\beq
z^n=\H{n}.
\eeq
In the same way it can be proved that
\beq
z^n\H{j}=\H{j+n}+\sum_{l=1}^nH^j_l\H{n-l}.
\eeq
Now it is enough to write this relation in the form
\beq
\H{j+n}-z^n\H{j}+\sum_{l=1}^nH^j_l\H{n-l}=0
\eeq
with the condition~\rref{receq}. 
\endpf
\subsection{The Quotient Spaces $\CQ_m$}
Let us consider now the second process of reduction. 
We concentrate on a vector field of the hierarchy, say 
$X_m$, and  we denote the corresponding time $t_m$ 
by
\beq
t_m=x,
\eeq
in order to point out its special role in the 
reduction.
This amounts to convert $t_m$ into a ``space  variable'', in the 
terminology used in the theory of KP equations.
It is worthwhile to remark that the projection to $\CQ_m$
with $m\neq 1$ formalizes the procedure that in the physics
literature~\cite{OlMa91}
is usually described as the ``exchange $x\leftrightarrow t_m$''.
\begin{defi} We call $\CQ_m$ the space of the solutions  of the 
$m$--th flow of the central system, i.e., the space of orbits of the 
vector field $X_m$.
\end{defi}
Thanks to the commutativity condition, any vector field 
$X_j$ of CS induces a flow, which we still denote by $X_j$, on 
$\CQ_m$. 
The first problem to be solved is to characterize the variety $\CQ_m$.
\begin{prop} 
The quotient spaces $\CQ_m$ 
can be identified with the space of 
$m$--tuples $\{H^{(a)}(z)\}_{a=1,\dots,m}$ of Laurent series of the form
\beq\label{generators}
H^{(a)}(z)=z^a+\sum_{l\geq 1}H^a_l(x)z^{-l},
\eeq
whose coefficients are functions of the space variable $x$.
\end{prop}
{\bf Proof.} 
It suffices to remark that the  
equations defining the vector field $X_m$
may be written in the form of recursion relations
\beq
\label{rrqm}
H^{(j+m)}=\left(\frac{\partial }{\partial x}+H^{(m)}\right)
H^{(j)}-\sum_{l=1}^jH^m_lH^{(j-l)}-\sum_{l=1}^mH^j_lH^{(m-l)}.
\eeq
They allow to compute the currents $(\H{m+1},\H{m+2},\dots)$ as 
differential polynomials in the first $m$ currents $(\H{1},\dots, 
\H{m})$. Explicitly, 
this means that the coefficients of $(\H{m+1},\H{m+2},\dots)$ 
are polynomials in the coefficients of $(\H{1},\dots,\H{m})$ and their 
$x$--derivatives.
Since the generators $(\H{1},\dots,\H{m})$ depend
on $x$ in a completely arbitrary way,
we conclude that the space of orbits
of the vector field $X_m$ coincides with the space
of $m$--tuples of Laurent series~\rref{generators}. 
\endpf
\section{Preliminary examples of reduced equations}
\label{sec4}
Let us consider the simplest examples of reductions of 
the central system.
We first consider the submanifold $\CS_2$. According to 
equation~\rref{receq}, it is parametrized by the
single Laurent series $\H{1}$.
To simplify the notations we set 
\beq\label{simplacca}
h(z):=\H{1}=z+\sum_{l\geq 1} h_lz^{-l}.
\eeq
The constraints defining $\CS_2$ are
\beq
\begin{array}{l}
\H{2}=z^2,\quad
\H{3}=z^2h(z)-h_1h(z)-h_2,\quad
\H{4}=z^4\\
\H{5}=z^2\H{3}-H^3_1h(z)-H^3_2\\
\phantom{\H{5}}=(z^4-h_1z^2+h_1^2-h_3)h(z)-h_2z^2+h_2h_1-h_4,
\end{array}
\eeq
and so on. 
To construct CS${}_2$, 
it is sufficient to plug these constraints in the equation 
\beq
\label{cs2eq}
\frac{\partial h(z)}{\partial t_j}=\H{j+1}-h\H{j}+\sum_{l=1}^jh_l
\H{j-l}+H^j_1.
\eeq
to get the equations displayed in Table~I.
\begin{table}[ht]
\hrule\medskip
\protect\caption{
The first CS${}_2$ equations}
\protect\[
\begin{array}{ll}
\dpt{h_1}{1}=-2h_2\qquad &\dpt{h_1}{3}=-2h_4+2h_1h_2\\
\dpt{h_2}{1}=-(2h_3+h_1^2)\qquad &\dpt{h_2}{3}=-2h_5+h_2^2+h_1^3\\
\dpt{h_3}{1}=-(2h_4+2h_1h_2)\qquad &\dpt{h_3}{3}=-2h_6+2h_1^2h_2\\
\dpt{h_4}{1}=-(2h_5+2h_1h_3+h_2^2)\qquad &\dpt{h_4}{3}=
-2h_7+2h_1^2h_3-h_3^2+h_1h_2^2\\
\dpt{h_5}{1}=-(2h_6+2h_1h_4+2h_2h_3)\qquad &
\dpt{h_5}{3}=-2h_8-2h_3h_4+2h_1^2h_4+2h_2h_2h_3\\
\end{array}
\protect\]
\protect\[
\begin{array}{l}
\dpt{h_1}{5}=2h_3h_2-2h_6+2h_1h_4-2(h_1)^{2}h_2\\
\dpt{h_2}{5}= h_3^{2}+2 h_2h_4+ h_1^{2}
h_3-2 h_7-h_1 h_2^{2}- (h_1 )^{4}\\
\dpt{h_3}{5}=2 h_3h_4-2h_1^{3}h_2-2 h_8+2
 h_1h_3h_2\\
\dpt{h_4}{5}=-2 h_9+3 h_1 h_3^{2}-2 
 h_1^{3}h_3+ h_4^{2}- h_2^{2} h_1^{2}+h_2^{2}h_3\\
\dpt{h_5}{5}= 2 h_2h_3^{2}-2 h_2 h_1^{2}
h_3-2 h_{10}+4 h_1h_3h_4-2 h_1^{3}h_4
\end{array}
\protect\]
\hrule
\bigskip
\end{table}
It is worthwhile to display also the first CS${}_3$
equations. 
By setting 
\beq
k(z):=\H{2}=z^2+\sum_{l\geq 1} k_lz^{-l}
\eeq
along with~\rref{simplacca}
and using the first parametric equations of the submanifold \CS${}_3$ 
\beq
\begin{array}{l}
\H{3}=z^3\\
\H{4}=z^3 h-h_1 k-h_2 h-h_3\\
\H{5}=z^3 k -k_1 k -k_2 h-k_3\\
\end{array}
\eeq
we get
the equations collected in Table~{II}.  
\begin{table}[ht]
\hrule\medskip
\protect\caption{The first CS${}_3$ equations}
\protect\[
\begin{array}{l}
{\dpt{h_1}{1}= {k_{{1}}-2\,h_{{2}}}}\\
{\dpt{h_2}{1}= {k_{{2}}-2\,h_{{3}}-{h_{{1}}}^{2}}}\\
{\dpt{h_3}{1}= {k_{{3}}-2\,h_{{2}}h_{{1}}-2\,h_{{4}}}{}}\\
{\dpt{h_4}{1}= {k_{{4}}-2\,h_{{1}}h_{{3}}-2\,h_{{5}}-{h_{{2}}}^{2}}}\\
\end{array}
\protect\]
\protect\[
\begin{array}{l}
{\dpt{h_1}{2}=\dpt{k_1}{1}= {-h_{{3}}-k_{{2}}+{h_{{1}}}^{2}}}\\
{\dpt{h_2}{2}=\dpt{k_2}{1}={-h_{{1}}k_{{1}}-h_{{4}}-k_{{3}}+h_{{2}}h_{{1}}}}\\
{\dpt{h_3}{2}=\dpt{k_3}{1}={-k_{{1}}h_{{2}}-h_{{1}}k_{{2}}-h_{{5}}-k_{{4}}+
        h_{{1}}h_{{3}}}}\\
{\dpt{h_4}{2}=\dpt{k_4}{1}={-k_{{1}}h_{{3}}-k_{{2}}h_{{2}}-h_{{1}}k_{{3}}-
        h_{{6}}-k_{{5}}+h_{{1}}h_{{4}}}}\\
\end{array}
\protect\]
\protect\[
\begin{array}{l}
{\dpt{k_1}{2}= {h_{{4}}+h_{{1}}k_{{1}}-h_{{2}}h_{{1}}-2\,k_{{3}}}}\\
{\dpt{k_2}{2}= {h_{{5}}-{h_{{2}}}^{2}-h_{{1}}k_{{2}}+
        2\,k_{{1}}h_{{2}}-2\,k_{{4}}-{k_{{1}}}^{2}}}\\
{\dpt{k_3}{2}= {h_{{6}}-h_{{1}}k_{{3}}-h_{{2}}h_{{3}}-
        2\,k_{{1}}k_{{2}}+2\,k_{{1}}h_{{3}}-2\,k_{{5}}}}\\
{\dpt{k_4}{2}={h_{{7}}-h_{{1}}k_{{4}}-h_{{2}}h_{{4}}-2\,k_{{1}}k_{{3}}+
        2\,h_{{4}}k_{{1}}-2\,k_{{6}}-{k_{{2}}}^{2}}}\\
\end{array}
\protect\]
\hrule
\bigskip
\end{table}
\par
Let us now consider the quotient space $\CQ_1$.
We keep the usual notation $\H{1}=h$,
but we recall that here $h$ must be considered as a  
a Laurent series $h=z+\sum_{l\ge1}h_l(x)/ z^{l}$, whose coefficients 
depend on the space variable $x$.
By using the equations of the vector field $X_1$, from~\rref{generators} 
we obtain the relations
\beq
\label{kp1con}
\begin{array}{l}
\H{2}=h_x+h^2-2h_1\\
\H{3}=\H{2}_x+h\H{2}-h_1h-h_2\\
\phantom{\H{3}}=h_{xx}+3hh_x+h^3-3h_1h-3(h_{1x}+h_2)\\
\cdots 
\end{array}
\eeq
The \kp{1} equations, describing the projection of 
CS on 
$\CQ_1$, are obtained by introducing the 
recursion relation~\rref{kp1con} 
in the equation
\beq
\label{kp1eq}
\frac{\partial h}{\partial t_j}=\H{j+1}-h\H{j}+\sum_{l=1}^jh_l
\H{j-l}+H^j_1=\frac{\del \H{j}}{\del x}.
\eeq
Expanding these equations in powers 
of $z$, one constructs the equations displayed in Table~{III}.
\begin{table}[ht]
\hrule\medskip
\protect\caption{The first \kp{1} equations}
\protect\[
\begin{array}{ll} 
\dpt{h_1}{1}=h_{1x}\qquad &\dpt{h_1}{2}=\partial_x(h_{1x}+2h_2)\\ 
\dpt{h_2}{1}=h_{2x}\qquad &\dpt{h_2}{2}=\partial_x(h_{2x}+h_1^2+2h_3)\\ 
\dpt{h_3}{1}=h_{3x}\qquad & \dpt{h_3}{2}=\partial_x(h_{3x}+2h_4+
2h_1h_2)\\
\dpt{h_1}{3}=\partial_x(h_{1xx}+3h_{2x}+3h_3) &\\ 
 \dpt{h_2}{3}=\partial_x(h_{2xx}+3h_{3x}+3h_1h_{1x}+3h_4+3h_1h_2) &\\ 
 \dpt{h_3}{3}=\partial_x(h_{3xx}+3h_{4x}+3h_1h_{2x}+3h_{1x}h_2
&\\
\phantom{ \dpt{h_3}{3}=}+3h_1h_3+3h_2^2+h_1^3) &
\end{array}
\protect\]
\hrule
\end{table}
After a suitable coordinate 
change~\cite{SS,Wil81,CFMP4}, 
these equations coincide with the usual KP equations in the
$1+1$ formalism. \par

We shall now consider the double reduction 
on the space $\CQ^1_2$.
We can start either from \kp{1} or from CS${}_2$. 
In the first case, we have only to
impose the constraint
\beq
\H{2}=h_x+h^2-2h_1=z^2
\eeq
that is,
\beq
\begin{array}{l}
2h_2+h_{1x}=0\\
2h_3+h_{2x}+h_1^2=0\\
2 h_4+h_{3x}+2 h_1 h_2=0\\
\cdots
\end{array}
\eeq
on the currents~\rref{kp1con}.
As it is well--known~\cite{Wil81}, this is the constraint 
defining the Hamiltonians of the KdV hierarchy, which
leads from the KP equations to the KdV equations 

In the second case, we can proceed
in three steps. First, 
we use the exactness condition~\rref{j-k}
to write the equations of Table~{I} in the form of 
Table~{IV}.
\begin{table}[ht]
\hrule\medskip
\protect \caption{ The $CS_2$ equations as conservation laws.}
\protect\[
\begin{array}{ll}
\dpt{h_1}{1}=-2h_2\qquad &\dpt{h_1}{3}=\dpt{}{1}(h_3-h_1^2)\\
\dpt{h_2}{1}=-(2h_3+h_1^2)\qquad &\dpt{h_2}{3}=\dpt{}{1}(h_4-h_1h_2)\\
\dpt{h_3}{1}=-(2h_4+2h_1h_2)\qquad &\dpt{h_3}{3}=\dpt{}{1}(h_5-h_1h_3)\\
\dpt{h_4}{1}=-(2h_5+2h_1h_3+h_2^2)\qquad &\dpt{h_4}{3}=
\dpt{}{1}(h_6-h_1h_4)\\
\dpt{h_5}{1}=-(2h_6+2h_1h_4+2h_2h_3)\qquad 
&\dpt{h_5}{3}=\dpt{}{1}(h_7-h_1h_5)\\
\end{array}
\protect\]
\protect\[
\begin{array}{l}
\dpt{h_1}{5}=\dpt{}{1}(h_5-2h_1h_3+h_1^3)\\
\dpt{h_2}{5}=\dpt{}{1}(h_6-h_2h_3-h_1h_4+h_2h_1^2)\\
\dpt{h_3}{5}=\dpt{}{1}(h_7-h_3^2-h_1h_5+h_3h_1^2)\\
\dpt{h_4}{5}=\dpt{}{1}(-h_3h_4+h_1^{2}h_4-h_1h_6+h_8)\\
\dpt{h_5}{5}=\dpt{}{1}(-h_3h_5-h_1h_7+h_9+h_1^{2}h_5)\\
\end{array}
\protect\]
\hrule
\end{table}
Then we transform the time $t_1$ in 
a space variable $x$, and we regard the first~\cs{2} equation 
as a set of recursive relations
\beq
\label{conkdv}
\begin{array}{l}
2h_2=-h_{1x}\\
2h_3=-(h_{2x}+h_1^2)\\
2h_4=-(h_{3x}+2h_1h_2)\\
\cdots
\end{array}
\eeq
They allow to identify the space $\CQ_2^1$
with the space of scalar functions $u\equiv 2 h_1$ in one
space variable $x$ (the phase space of KdV). 
Finally, we insert these constraints in the first 
component of each vector field $X_j$ of~\cs{2} (the other components 
can be neglected, since they give rise only to differential 
consequences of the previous ones).
Once again, we get the KdV equations
\beq
\begin{array}{l}
\dpt{u}{3}=\del_x(h_3-h_1^2)=\frac18 \del_x(u_{xx}-3u^2)\\
\dpt{u}{5}=\del_x(h_5-2h_1h_3+h_1^3)=\frac1{32} 
\del_x(u_{xxxx}-10uu_{xx}-5u_x^2+10u^3)\\
\end{array}
\eeq
and so on.
This computation clearly points out how the projection process
allows to transform a family of dynamical system 
into a hierarchy of  evolution partial differential equation in $1+1$ 
dimensions (one space--dimension and one time--dimension).\par

We end this section by briefly describing 
another example of double 
 reduction of  CS,
on the intersection $\CS_n\cap \CZ_p$. In this case
we obtain
systems of ordinary differential equations in a finite
number of fields.
Let us exemplify this feature by considering
the intersection $\CS_2\cap \CZ_5$.
Looking at the equations defining $X_5$ in Table~{III}, 
we obtain the constraints
\beq
\label{concs25}
\begin{array}{l}
h_6=h_1h_4+h_2h_3-h_1^2h_2\\
h_7=h_2h_4+\frac12(h_3^2+h_1^2h_3-h_1h_2^2-h_1^4) \\
h_8=h_3h_4+h_1h_2h_3-h_1^3h_2\\
\cdots
\end{array}
\eeq
which 
are an infinite system of recurrence 
relations allowing to express all the $h_l$'s, for $l\ge 6$, as 
polynomial functions of $h_1,\dots,h_5$. Therefore the  invariant 
submanifold $\CS_2\cap \CZ_5$
is 5--dimensional . The restriction of \cs{2} to this  
submanifold is simply constructed by plugging the constraints 
\rref{concs25} in the first five components of the vector fields $X_1$ 
and $X_3$. We get the equations
\beq
\label{eqcs25}
\begin{array}{ll}
\dpt{h_1}{1}=-2h_2 & \dpt{h_1}{3}=-(2h_4-2h_1h_2)\\
\dpt{h_2}{1}=-(2h_3+h_1^2) & \dpt{h_2}{3}=-(2h_5-h_2^2-h_1^3)\\
\dpt{h_3}{1}=-(2h_4+2h_1h_2)& \dpt{h_3}{3}=-(2h_1h_4+2h_2h_3-4h_1^2h_2)\\
\dpt{h_4}{1}=-(2h_5+2h_1h_3+h_2^2)& 
\dpt{h_4}{3}=-(2h_3^2-h_1^2h_3-h_1^4-2h_1h_2^2+2h_2h_4)\\
\dpt{h_5}{1}=-(4h_1h_4+4h_2h_3-2h_1^2h_2)&
\dpt{h_5}{3}=-(4h_3h_4-2h_1^3h_2-2h_1^2h_4)
\end{array}
\eeq
The equations corresponding to the vector fields $X_{2j+1}$, for $j\ge 
2$, are combinations of equations \rref{eqcs25}.
Hence the equations we have written represent the whole reduced 
system.  
After a suitable coordinates change, 
they coincide with the second Novikov system~\cite{BN} 
relative to the KdV hierarchy.\par

\section{The equations \cs{n}  and \kp{m} }
\label{sec5}
In Section~\ref{sec3} we explained the geometric process
leading to  the \cs{n} and \kp{m} equations.
In this section we  shall give an algebraic definition
of these equations by using
the 
concept of {\em \Fdb\ basis\/} associated with a (finite set of) monic 
Laurent series. We present this idea first for the \cs{n} case. In 
Section 3 we showed that this is a  system of ordinary differential 
equations in the first $(n-1)$ currents
\beq
\H{a}=z^a+\sum_{l\ge 1} H_l^a z^{-l},
\eeq
for $a=1,\dots,n-1$. As usual, we set $\H{0}=1$ and we 
we consider the point $\H{n}=z^n$ in the space 
$\CL$ of all truncated Laurent series. We 
call {\em stationary \Fdb\ basis} at the point $z^n$ the basis of $\CL$ 
defined by the iterates
\beq
\begin{array}{l}
\F{j+n}=z^n\cdot\F{j}\\
\F{a}=\H{a}\qquad\mbox{for }a=0,\dots,n-1,
\end{array}
\eeq
of the initial generators $(\H{0},\dots,\H{n-1})$, for $j\in\Bbb Z$. 
Let us denote with $\CHp$ the subspace of $\CL$ spanned by the 
nonnegative elements of the \Fdb\ basis at the point $z^n$, 
and with $\CHm$ the subspace 
spanned by the negative ones. We decompose $\CL$ in the direct sum
\beq
\label{decomplau}
\CL={\CH}_+\oplus {\CH}_-,
\eeq
and we call 
\beq
\label{projzj}
\H{j}=\pi_+(z^j)
\eeq
the projections on ${\CH}_+$ of the powers $z^j$ with respect to this
decomposition.
\begin{defi} {\bf (Second definition of \cs{n})}
We call \cs{n} the dynamical system in the currents
$(\H{1},\dots,\H{n-1})$ defined by
\beq
\label{csneqfdb}
\dpt{\H{a}}{j}=-\pi_-(\H{a}\H{j}),
\eeq
for $a=1,\dots,n-1$, and $j\in\Bbb N$.
\end{defi}\par
We proceed in the same way for the \kp{m} equations. In this case
$\CL$ is the space of truncated Laurent 
series whose coefficients are functions of the  space variable $x$, and 
we consider the point
\beq
\H{m}=z^m+\sum_{l\ge 1} H_l^m z^{-l}.
\eeq
We call {\em differential \Fdb\ basis} at the point $\H{m}$ 
the basis of $\CL$ defined by the iterates
\beq
\label{fdbdif}
\begin{array}{l}
\F{j+m}=(\del_x+\H{m})\cdot\F{j}\\
\F{a}=\H{a}\qquad\mbox{for }a=0,\dots,m-1,
\end{array}
\eeq
of the initial generators $(\H{0},\dots,\H{m-1})$, for $j\in\Bbb Z$. 
We observe that the computation of the elements $\H{j}$, with $j<0$, 
of the \Fdb\ basis does not require any integration, since the 
recurrence relation \rref{fdbdif} can be inverted in a purely algebraic 
way. As before, we decompose the space $\CL$ in the form 
\rref{decomplau}, and we introduce the projections 
\beq
\label{projzja}
\H{j}=\pi_+(z^j)
\eeq
of the powers $z^j$. Even if we used the same 
notations to point out the analogy between the two situations, one has 
to keep in mind that they live in different spaces, and that the \Fdb\ 
basis are defined in a different way.
\begin{defi}  {\bf (Second definition of \kp{m})}
We call \kp{m} the system of evolutionary partial differential equations 
in the currents $(\H{1},\dots,\H{m})$ defined by
\beq
\label{kpmeqfdb}
\dpt{\H{a}}{j}=-\pi_-(\H{a}\H{j}),
\eeq
for $a=1,\dots,m$, and $j\in\Bbb N$.
\end{defi}
Now we have to show that the equations defined above  coincide with the 
ones obtained by the reduction process.
\begin{prop}\label{equi}
The equations \rref{csneqfdb} and \rref{kpmeqfdb} constructed by the 
\Fdb\ basis algorithm coincide with the reduced equations \cs{n} and \kp{m} 
defined by  the reduction scheme.
\end{prop}
{\bf Proof.} We call ${\widehat{\CH}}_+$ the subspace of $\CL$ 
spanned by the currents $\H{j}$ of the central system (the same 
subspace that we denoted with $\CHp$ in Section 3). 
We consider ${\widehat{\CH}}_+$ at a 
generic point of the submanifold $\CS_n$. From \rref{zinv}, at the 
points of $\CS_n$ the subspace ${\widehat{\CH}}_+$ is invariant with 
respect to multiplication by $z^n$. Hence it contains all nonnegative 
elements of the stationary \Fdb\ basis associated with the point 
$z^n$. Therefore ${\widehat{\CH}}_+$ at the points of $\CS_n$ 
coincides with the subspace $\CHp$ appearing in the decomposition
\rref{decomplau}, constructed by choosing as generators the currents 
$(\H{1},\dots,\H{n-1})$ defining the point of $\CS_n$. It follows that 
the projection \rref{projzj} belongs to ${\widehat{\CH}}_+$, and 
therefore coincides with the 
corresponding current of CS evaluated at $\CS_n$. The first part of 
the proposition then follows from Proposition \ref{propforcs}.\par
As far as the \kp{m} equations are concerned, one can argue in the 
same way. In this case we evaluate the subspace ${\widehat{\CH}}_+$ at 
the points of a generic integral curve of the vector field $X_m$. We 
use the invariance property \rref{invrel} to conclude that at these 
points the subspace ${\widehat{\CH}}_+$ contains all nonnegative 
elements of the differential \Fdb\ basis \rref{fdbdif}, and therefore 
it coincides with the subspace $\CHp$ associated with the point 
$\H{m}$. This observation leads to the conclusion that the projection 
\rref{projzja} coincides with the current $\H{j}$ of CS evaluated at 
the points of the integral curve of the vector field $X_m$. This 
suffices to prove the coincidence between the \kp{m} equations 
obtained by projection and equations \rref{kpmeqfdb}.
\endpf
To enlighten the (differential) 
\Fdb\  algorithm, we consider the 
\kp{2} equations. We put $h=\H{1}$ and $k=\H{2}$. The first elements 
of the \Fdb\ basis at the point $k$, associated  with the 
generators $(1,h)$,  are
\beq
\begin{array}{ll}
\F{0}=1&\quad\F{3}=h_x+kh\\
\F{1}=h&\quad\F{4}=k_x+k^2\\
\F{2}=k&\quad\F{5}=h_{xx}+2kh_x+hk_x+k^2h.
\end{array}
\eeq
We combine these elements in the form
\beq
\label{curkp2}
\begin{array}{ll}
\H{1}=\F{1}&\quad\H{3}=\F{3}-h_1\F{1}-(h_2+k_1)\F{0}\\
\H{2}=\F{2}&\quad\H{4}=\F{4}-2k_1\F{1}-2k_2\F{0}
\end{array}
\eeq
in order to obtain Laurent series with the asymptotic behavior $\H{j}
=z^j+O(z^{-1})$. They are the currents $\H{j}$ written as differential 
polynomials of $h$ and $k$. To obtain the \kp{2} equations we can 
either compute the projections $\pi_-(h\H{j})$ and $\pi_-(k\H{j})$, or 
write the equations 
\beq
\begin{array}{l}
\label{syskp2}
\dpt{}{j}h=\H{j+1}-h\H{j}+\sum_{l=1}^jh_l\H{j-l}+H^j_1\\
\dpt{}{j}k= \del_x \H{j}.
\end{array}
\eeq
In particular, the first vector field of the \kp{2} hierarchy is given by
\beq
\label{kp2eq1}
\begin{array}{l}
\dpt{}{1}h=k-h^2+ 2 h_1\\
\dpt{}{1}k=\del_x h.
\end{array}
\eeq
This computation allows to
illustrate the link between \kp{1} and \kp{2}. 
Indeed we write equations~\rref{kp2eq1} 
in the form
\beq
\label{kp2eq1b}
\begin{array}{l}
k=\dpt{}{1}h+h^2- 2 h_1\\
\del_x h=\dpt{}{1}k
\end{array}
\eeq
and we take the $x$--derivative of the first equation above to get
\[
\del_x k=\dpt{}{1}h_x+ 2 h h_x -2 h_{1x}.
\]
By substitution we get 
\beq
\label{kx}
\del_x k=\frac{\del^2}{\del t_1^2} k+ 2 h \dpt{}{1}k 
- 2\dpt{}{1}k_1.
\eeq
By this process, we can express  $h$, $k$, and the collection of 
their ``space derivatives'' $(h_x, k_x ; h_{xx} , k_{xx} ; \dots)$ as 
$t_1$--differential polynomials in $h$.
As a consequence, all the currents $\H{j}$
are expressed as  $t_1$--differential 
polynomials in $h$, and the \kp{2}--system \rref{syskp2} reduces to 
the \kp{1} equations with $x=t_1$,
\beq
\dpt{}{j}h=\dpt{}{1}\H{j}.
\eeq
Conversely, the \kp{2} \ger y can be obtained as a differential 
prolongation of \kp{1}. More precisely, 
starting from ~\kp{1}, 
one has to put $k:=\H{2}$ and to 
use the definition of $k$ and the second equation of \kp{1},
\beq
\begin{array}{l}
\label{kp2-kp1}
k=h_x+h^2- 2 h_1\\
\dpt{}{2}h=k_x
\end{array}
\eeq
to express all the $x$--derivatives of $h$ in terms of $h$, $k$ and their 
$t_2$--derivatives. 
At this point one simply changes the names of the 
variables $x$ and $t_2$, by putting at first $x\mapsto t_1$ and then 
$t_2\mapsto x$. 
This kind of procedure is easily extended to describe
the relations between the \kp{1} and \kp{m} equations,
for $m\ge 3$.
As a matter of fact, the existence of such a link between different \kp{m}
theories,
which gives rise to these mappings of flows,
is extremely natural in our approach. It is a consequence of the
fact that all \kp{m}
theories are {\em different} field--theoretic pictures 
of the {\em same} infinite 
dimensional dynamical system, the central system.
\section{The \kdv{m}{n} equation}\label{sec6}
The \kdv{m}{n} 
equations are the restriction of the \kp{m} equations to the 
invariant submanifold $\CQ_n^m$ of the orbits of $X_m$ which are tangent 
to $\CS_n$ (see Figure~1). In the case $n> m$, 
the space $\qmn{m}{n}$ is characterized by the 
following property.
\begin{prop}
Let $(\H{1},\ldots,\H{m})$ be a generic point 
of $\qmn{m}{}$, and let \hfill\linebreak
$(\H{1},\ldots,\H{n-1})$ be the  
first $(n-1)$ currents  
constructed by the (differential)
\Fdb\ algorithm of Section~\ref{sec5}. The point $(\H{1},\dots,\H{m})$
belongs to $\qmn{m}{n}$ if and only if the Laurent series $(\del_x+\H{m})
\H{\al}$, for $\al=0,\ldots, n-1$ admit the linear expansion
\beq\label{frob1}
(\del_x+\H{m})\H{\al}=\sum_{\be=0}^{n-1}u_\be^\al(\la) \H{\be}
\eeq
where the coefficients $u_\be^\al(\la) $ 
depend polynomially on $\la=z^n$
and on the coordinates $H^1_l,\ldots, 
H^m_l$ of the point of $\qmn{m}{n}$.
\end{prop}
{\bf Proof.} We recall that the vector field $X_m$ is defined by the equation
\beq
\left(\dpt{}{m}+\H{m}\right)(\CHp)\subset \CHp
\eeq
and that $\CS_n$ is characterized by the property
\beq 
z^n\cdot(\CHp)\subset\CHp.
\eeq
Thus, on $\CS_n$ the subspace $\CHp$ is generated by 
$(\H{1},\dots,\H{n-1})$ and by multiplication by powers of $\la$. 
Equation~\rref{frob1} simply means that $X_m$ is tangent to $\CS_n$.
\endpf
To see how this condition works in practice, we will restrict from now 
on to the case of \kdv{2}{3}. In the notations of Section~\ref{sec4},
i.e., $h(z)=\H{1},k(z)=\H{2}, \la=z^3$, 
the submanifold $\qmn{2}{3}$ is defined by the constraints
\beq\label{frcon}\label{kp23}
(\del_x+k(z))
\left[ \begin{array}{c} 
1\\
h(z)\\
k(z)
\end{array}\right]                                    
=\CU(\la)
\left[ \begin{array}{c} 
1\\
h(z)\\
k(z)
\end{array}\right]. 
\eeq
The matrix $\CU$ 
is easily computed by comparing the powers of $z$
on both sides. We obtain the {\em generalized Frobenius matrix}
\beq\label{frobmat}
\CU=\left(
\begin{array}{ccc}
0 & 0 &1\\
\la+h_2+h_1 & h_1 & 0\\
2 k_2-h_3 & \la + 2k_1-h_2& -h_1
\end{array}
\right).
\eeq
Therefore the constraints~\rref{frcon} are equivalent to the pair
of Riccati equations
\bea
&&h(z)_x+h(z)k(z)=z^3+h_1h(z)+(h_2+k_1)\label{uno23}\\
&&k(z)_x+k(z)^2=-h_1k(z)+h(z)(2k_1-h_2+z^3)-(h_3-2 k_2).\label{due23}
\eea
\begin{prop}\label{finitefield}
The $\qmn{2}{3}$ constraints allow to express all the fields 
$\{h_j,k_j\}_{j\geq 3}$ of the \kp{2} theory as differential polynomials 
of the first four components $h_1,h_2,k_1,k_2$. Therefore the space
$\qmn{2}{3}$ can be identified with the space of generalized Frobenius 
matrices~\rref{frobmat} and \kdv{2}{3} reduces to
a hierarchy on four fields, $h_1,h_2,k_1,k_2$. 
\end{prop}
{\bf Proof.} 
The coefficient of $z^{-1}$ of~\rref{uno23} gives
\beq
\label{acca323}
h_3=h_1^2-h_{1x}-k_2,
\eeq
and, in general, the coefficients of $z^{-i}$ of~\rref{due23} and of 
$z^{-(i+1)}$ of~\rref{uno23} can be read as a linear system in the
unknowns $(k_{i+2},h_{i+3})$ of the form
\beq
\left[\begin{array}{cc} 2 & -1\\ 1 & 1 \end{array}\right]
\left[\begin{array}{c} k_{i+2}\\ h_{i+3}\end{array}\right]
=\left[\begin{array}{c} \al_i\\ \be_{i}\end{array}\right]
\eeq
where the RHS is a $x$--differential polynomial in the Laurent coefficients 
$\{h_l,k_m\}$ with $l< i+3,m< i+2$.
\endpf

To obtain the \kdv{2}{3} hierarchy 
in the form of a zero--curvature equation 
we consider the
vector field $X_j$ and the associated current $\H{j}$ constructed with 
the differential \Fdb\ algorithm of Section~\ref{sec5}. 
Since this vector field is tangent to $\CS_3$ 
at the points of $\qmn{2}{3}$, there exists
a matrix $\V{j}(\la)$, whose entries 
$\v{j}_\be^\al(\la)$ are polynomials in $\la=z^3$
with coefficients depending on the Laurent coefficients $\{h_i\}$ and
$\{k_i\}$, such that
\beq\label{frobt}
\left(\dpt{}{j}+\H{j}\right)\H{\al}
=\sum_{\be=0}^{n-1}\v{j}_\be^\al(\la) \H{\be}
\eeq
for $\al=0,1,2$.
\begin{prop}
The \kdv{2}{3} \ger y on the space of generalized 
Frobenius matrices~\rref{frobmat}
admits the zero--curvature representation
\beq\label{zaksha}
\dpt{}{j}\CU-\frac{\del}{\del x}\V{j}+[\CU,\V{j}]=0.
\eeq
\end{prop}
{\bf Proof.} It is enough to cross differentiate equations~\rref{frob1} and
\rref{frobt} defining the matrices $\CU$ and $\V{j}$, and to recall
the result of Proposition~\ref{finitefield} about the possibility of 
expressing the Laurent coefficients $h_j, k_j,\quad j\ge 3$, in terms 
$(h_1,h_2,k_1,k_2)$.
\endpf
The
same procedure can be used for other fractional KdV equations.
It
should be compared with the approach of~\cite{Hol1,Hol2}.

\section{On the Bihamiltonian aspects of \kdv{2}{3} }\label{sec7}
To show that our \kdv{2}{3} hierarchy coincides with the 
fractional  ${\sl23}$ KdV hierarchy of \cite{OlMa91}, 
it could suffice to remark that \kdv{2}{3} can be also 
obtained from \kdv{1}{3} (which is the classical
Boussinesq hierarchy) by means of the 
$x\leftrightarrow t_2$ interchange, in the same 
way as  \kp{2} is obtained from \kp{1}.  
Actually, this hierarchy was first 
studied in~\cite{BaDe91} by means of the Hamiltonian 
reduction suggested in~\cite{Po90,Be91}, and the existence 
of two compatible Hamiltonian structures was also pointed out.
The theory
of fractional KdV hierarchy has been further explored and generalized
in a number of papers (see, e.g., \cite{Hol1,Hol2,FeHaMa93,FGMS95}),
by means of the study of its Lie--algebraic aspects.\par
In this section we will briefly address the problem of
showing a direct relation between the \kdv{2}{3} treated in 
Section~\ref{sec6} and the corresponding generalized $\sl23$
DS one of \cite{BaDe91,Hol1}, assuming as a 
starting point its \bih\ aspects. 
Full proofs and detailed explanations will be given elsewhere~\cite{CFMP8}.
The reader should keep in mind the logical path followed in 
Section~\ref{sec2}, where the \kp{1} hierarchy  has been 
derived from the Poisson pencil of KdV.\par
We recall that a manifold $\CM$ is said to be \bih\ if it is endowed with 
two compatible \parp s,
and that a \parp\ can be assigned in 
terms of a \tenp\ $P:T^*\CM\to T\CM$ as
\beq
\label{tenp}
\{f,g\}=\langle df,Pdg\rangle.
\eeq
We consider the Lie algebra ${\fraksl}(3)$ and its 
loop algebra $\Alg=L({\frak sl}(3))$, i.e., the space of 
$C^\infty$--maps from $S^1$ to $\fraksl(3)$. 
The algebra $\Alg$ is a   
\bih\ manifold (see, e.g.,~\cite{DS,CMP}).
The Poisson structures
we will consider hereinafter are:
\bea
\label{poisM}
(P_0)_M\cdot V &=& [A_2,V]\\
(P_1)_M\cdot V&=& V_x+[V,M].
\eea
Here, $M$ is a point in $\Alg$, $V$ is a cotangent vector at $M$,
$x$ is the parameter on $S^1$, and $A_2$ is the element
\beq
\label{A}
A_2=\left(\begin{array}{ccc}
0 & 0 & 0\\
1 & 0 & 0\\
0 & 1 & 0
\end{array}\right)
\eeq
in $\Alg$.
To lower the number of degrees of freedom,
one seeks (following the seminal paper~\cite{DS}) 
a Poisson reduction of this bihamiltonian structure.
According to the \bih\  variant~\cite{CMP} of the Marsden--Ratiu
reduction theorem~\cite{MR}, one has to
consider a symplectic leaf $\CS$  of $P_0$, the (integrable) distribution $D=
P_1(\ker P_0)$, and the intersection $E=D\cap T\CS$. Then 
the quotient manifold $\CN=\CS/E$ is still a \varb. 
In our case,  we choose $\CS$ to be the symplectic leaf 
passing through the point
\beq
\label{B}
B_2=\left(\begin{array}{ccc}
0 & 0 & 1\\
0 & 0 & 0\\
0 & 0 & 0
\end{array}\right).
\eeq
The submanifold  $\CS$ has dimension $6$ 
over $C^\infty(S^1)$, and its generic 
point can be parametrized as
\beq
\label{S}
S=\left(\begin{array}{ccc}
p_0(x) &   p_2(x)    & 1\\
q_0(x) &   p_1(x)-p_0(x) & -p_2(x)\\
q_2(x) &   q_1(x) &-p_1(x)
\end{array}\right).
\eeq
Then we consider the distribution $D$, which turns out to be tangent to 
$\CS$, so $E=D$. 
The quotient manifold $\CN=\CS/E$ is parametrized by the four fields
\beq\label{ufields}
\begin{array}{lll}
u_0&=&-p_0+p_1+p^2_2\\ 
u_1&=&q_0+ 2 p_0 p_2- p_1 p_2 -p_2^3+p_{2 x}\\
u_2&=&q_1 - p_0 p_2+2 p_1 p_2 +p_2^3+p_{2 x}\\
u_3&=&q_2 +p_{0x} -p_2 p_{2x} + q_0 p_2 -q_1 p_2 + p_0 p_1 - p_2^4 - 
2 p_1 p_2^2.
\end{array}
\eeq
The reduced Poisson pencil $P_\la^\CN=P_1^\CN-\la P_0^\CN$ on $\CN$ 
turns out to be 
\beq
\label{poisN}
P_\la^\CN=\left(\begin{array}{cccc}
\frac{2}{3}\del_x       & - u_1-\la             & u_2+\la &\left( 
\frac{1}{3}\del_x^2 - u_0\del_x -u_{0x}\right)\\
 \ast & 0 &(P_\la^\CN)^{23}& 
                                        \left(2 u_1\del_x 
                                         +u_{1x}- 2 u_0 u_1\right)
                                         -\la\left(-2\del_x+2u_0\right)\\
 \ast & \ast & 0 & \left(u_2\del_x
+ u_{2x} + 2 u_0 u_2\right)-\la\left(-2\del_x-2u_0\right)\\
 \ast & \ast & \ast &
(P_\la^\CN)^{44}
\end{array}\right),
\eeq
where
\[
\begin{array}{l}
(P_\la^\CN)^{23}=-\del_x^2 + 3 u_0 \del_x
                        + 2 u_{0x}+ u_3 
                       - 2 u_0^2\\ 
(P_\la^\CN)^{44}=
- \frac{2}{3} \del_x^3 + \left(\frac{4}{3} u_{0x}+ 2 u_3
+ \frac{2}{3} u_0^2\right)\del_x
                + \frac{2}{3} u_0 u_{0x}+  u_{3x}
               + \frac{2}{3} u_{0xx}.
\end{array}
\]
They are computed according to the standard procedure explained 
in~\cite{MR,CMP}.
The reduced Poisson tensor $P_1^\CN$ coincides with the one given in 
\cite{Hol2}, after the change of coordinates
\beq
\label{cooHol}
u_0=-\tilde U,\quad u_1={\tilde G}^+,\quad
u_2={\tilde G}^-,\quad u_3=\tilde T-{\tilde U}^2+{1\over 2}{\tilde 
U}_x
\eeq
and $x\mapsto -x$.\par

The basic remark to connect this theory with the \kdv{2}{3} treated in 
Section~\ref{sec6} is that the quotient space $\CN$ can be identified
with the space of (generalized) Frobenius matrices
\beq\label{cu}
\CU=\left(
\begin{array}{ccc}
0 & 0 & 1\\
u_1& u_0& 0\\
u_3& u_2 & -u_0
\end{array}
\right)
\eeq
by noticing that the constraints
$p_0=p_2=0$ define a submanifold $U$ of $\CS$ {\em transversal} to the 
distribution $D$. Consequently, by comparing the matrices~\rref{frobmat}
and~\rref{cu} we arrive at the identification
\beq\begin{array}{c}
u_0=h_1,\quad
u_1=h_2+k_1,\quad
u_2=2k_1-h_2\\
u_3=2k_2-h_3=3k_2-h_1^2+h_{1x},
\end{array}
\eeq
which can be inverted in local form to yield 
\beq
\begin{array}{l}
h_1=u_0,\qquad
h_2=\frac{1}{3}(2 u_1 -u_2)\\ 
k_1=\frac{1}{3}(u_1+u_2),\quad 
k_2=\frac{1}{3}(u_3+u_0^2-u_{0x}).
\end{array}
\eeq
This procedure sets up a diffeomorphism
\beq
\Phi:\CN\longrightarrow \qmn{2}{3}
\eeq
between the reduced bihamiltonian manifold $\CN$ and the phase space of the
\kdv{2}{3} theory.
This diffeomorphism enjoys the following properties, which we limit 
ourselves to 
state without proofs (they will be detailed elsewhere~\cite{CFMP8}):
\begin{enumerate}
\item
The integral on $S^1$ 
\beq
H(z)=3z^2\int_{S^1}k(z)\,dx,
\eeq
of the pull--back of the second current
of the \kdv{2}{3} theory
is a Casimir function of the Poisson pencil~\rref{poisN}.
\item
The push--forward of the
Hamiltonian vector fields on $\CN$ associated with this Casimir
function are
the \kdv{2}{3} flows~\rref{zaksha} on $\qmn{2}{3}$. Hence, we
are allowed to identify the two theories.
\end{enumerate}
This result can be usefully compared with the KdV case briefly treated in
Section~\ref{sec2}. The KdV (=\kdv{1}{2}) theory is a theory in a single field
$u$, defined on the quotient space $\CN_2^1$ associated with 
the Lie--Poisson pencil on the loop algebra of $\fraksl_2$ and with the 
matrices~\cite{CMP}
\begin{equation}
A_1=\left(\begin{array}{cc}
0 & 0\\
1 & 0
\end{array}\right)
\qquad B_1 =\left(\begin{array}{cc}
0 & 1\\
0 & 0
\end{array}\right).
\end{equation}
The \kdv{2}{3} theory is a theory  on four fields, $(u_0,u_1,u_2,u_3)$ defined
on the quotient space $\CN_3^2$ associated with 
the Lie--Poisson pencil on the loop algebra of $\fraksl_3$ and with the 
matrices
\begin{equation}
A_2=\left(\begin{array}{ccc}
0 & 0 & 0\\
1 & 0 & 0 \\
0 & 1 & 0 
\end{array}\right)
\qquad B_2 =\left(\begin{array}{ccc}
0 & 0 & 1\\
0 & 0 & 0 \\
0 & 0 & 0 
\end{array}\right).
\end{equation}
In the first case the Casimir function of the reduced Poisson pencil
is defined by the solution $h(z)$ of the {\em single} Riccati 
equation
\beq\label{amoric}
h_x+h^2=u+z^2.
\eeq
In the latter case, the Casimir function is computed by solving the {\em pair}
of Riccati equations
\beq \begin{array}{ll}
h_x+kh&=u_0 h+(u_1+\la)\\
k_x+k^2&=-u_0k+(u_2+\la)h+u_3,
\end{array}
\label{5.21}
\eeq
in the first two currents $\H{1}=h(z)$, $\H{2}=k(z)$.
The appearance of these
systems of Riccati equations is a general feature of \kdv{n}{m} theories
with $m\ge 2$ which, in our opinion, deserves further attention.\par
When the point $u$ evolves according to the KdV hierarchy, the solution $h$
of the Riccati equation~\rref{amoric} evolves according to the
\kp{1} equations
\beq
\dpt{}{j} h =-\pi_-(h\H{j})
\eeq  
defined in Section~\ref{sec2}. Similarly, when the point
$(u_0,u_1,u_2,u_3)$ evolves according to the \kdv{2}{3} hierarchy,
the solution $(h,k)$ of the Riccati system~\rref{5.21}
evolves according to the \kp{2} equations
\beq\begin{array}{cc}
\dpt{}{j} h &=-\pi_-(h\H{j})\\
\dpt{}{j} k &=-\pi_-(k\H{j})
\end{array}.
\eeq  
In both cases the current densities $\H{j}$, which are constructed
in a different way in the two theories evolve in time according to the 
central system~\rref{00a}.
This completes the view of the relations connecting the central system
and the fractional KdV hierarchies we discussed in the case of \kdv{2}{3}.

\end{document}